\documentclass[twocolumn,amsmath,amssymb]{snp}
\usepackage{graphicx}
\usepackage{dcolumn}
\usepackage{bm}
\begin{document}

\title{{\Large Mass Spectra of $D$, $D_s$ Mesons using Dirac formalism with martin-like confinement potential}}

\author{\large Manan Shah$^1$}\email{mnshah09@gmail.com}
\author{\large Bhavin Patel$^2$}\email{azadpatel2003@yahoo.co.in}
\author{\large P C Vinodkumar$^1$}\email{p.c.vinodkumar@gmail.com}
\affiliation{$^1$Department of Physics, Sardar Patel University,Vallabh Vidyanagar-388120, INDIA}
\affiliation{$^2$P. D. Patel Institute of Applied Sciences, CHARUSAT, Changa-388421, INDIA}

\maketitle

\section*{Introduction}
Remarkable progress at the experimental side, with various high energy machines such as BaBar, BELLE, B-factories, Tevatron, ARGUS collaborations,
CLEO, CDF, D${\O}$ etc., for the study of hadrons has opened up new challenges in the theoretical understanding of light-heavy flavour hadrons.
The existing results on excited heavy-light mesons are partially inconclusive, and even contradictory in several cases. The predictions of masses of heavy-light system for low-lying 1S and $1P_J$ states of D and $D_s$ mesons were known from experiment \cite{PDG2012} and few from the theory \cite{Godfrey,Eichen,Ebert}. Here we study the mass spectra of $D$ and $D_s$ mesons in a relativistic framework.\\

\section*{Theoretical Framework}
 The bound constituent quark and antiquark inside the meson are in definite energy states having no definite momenta. However one can find out the momentum distribution amplitude for the constituent quark and antiquark inside the meson immediately before their annihilation to a lepton pair. Though the colour confinement of quarks are understood in terms of multigluon exchanging at the non-perturbative regime of the hadronic size, it is not feasible to compute theoretically from the QCD first principles. Thus one assumes various confinement mechanism to study the hadronic properties. In the present study, we assume that the constituent quarks in a meson core is independently confined by an average Martin-like potential of the form \cite{Barik2000}
\begin{equation}\label{eq:a}
V(r)= \frac{1}{2} (1+\gamma_0) (\lambda r^{0.1}+V_0)
\end{equation}

To a first approximation, the confining part of the interaction is believed to provide the zeroth-order quark dynamics inside the meson core through the quark Lagrangian density
\begin{table}[b]
\begin{center}
\tabcolsep 0.3pt
    \small
\caption{S-wave $D$ ($c\bar{q}$, q = d,u) spectrum (in MeV).} \label{tab1}
\begin{tabular}{|c|c|c|c|c|c|c|}
\hline
  &      &           & \multicolumn{1}{c} {Experiment} && &\\
\cline{4-5}
nL & State &  Present  &  Meson  & Mass \cite{PDG2012} & \cite{Badalian2011} & \cite{Ebert2010}\\
\hline
1S   &$1{^3S_1}$&  2013.3 & $D^*$(2010) &  2010.28$\pm$0.13 & - & 2010 \\
     &$1{^1S_0}$&  1874.0 &             &  1869.62$\pm$0.15 & - & 1871 \\

2S   &$2{^3S_1}$& 2581.0 &  $D^*$(2600) &  2608.7 \cite{del} & 2639 & 2632\\
     &$2{^1S_0}$& 2501.7 &  $D$(2550)  &  2539.4 \cite{del} & 2567 & 2581 \\

3S   &$3{^3S_1}$& 3088.9 &              &  -    & 3125 & 3096 \\
     &$3{^1S_0}$& 3031.5 &              &  -    & 3065 & 3062 \\

4S   &$4{^3S_1}$& 3567.8 &              &  -    & -    & 3482 \\
     &$4{^1S_0}$& 3521.6 &              &  -    & -    & 3452 \\
\hline
1P &$1{^3P_2}$&  2455.1 & $D_{2}^*$(2460) &  2462.6 $\pm$ 0.7     & - & 2460 \\
   &$1{^3P_1}$&  2348.0 &                 &      -                 & - & 2469 \\
   &$1{^3P_0}$&  2276.6 & $D_{0}^*$(2400) &  2318 $\pm$ 29        & - & 2406 \\
   &$1{^1P_1}$&  2317.3 & $D_{1}$(2420)   &  2421.3 $\pm$ 0.6     & - & 2426 \\

2P  &$2{^3P_2}$&  2907.0 &                  &   -      & 2965 & 3012 \\
    &$2{^3P_1}$&  2834.4 &                  &   -      & 2960 & 3021 \\
    &$2{^3P_0}$&  2786.0 &                  &    -     & 2880 & 2919 \\
    &$2{^1P_1}$&  2812.3 &                  &     -    & 2940 & 2932 \\
    \hline
\end{tabular}
\end{center}
\end{table}
\begin{equation}\label{eq:b}
{\cal L}^0_q (x)= \bar{\psi}_q(x) \left[\frac{i}{2} \gamma^{\mu} \overleftrightarrow{\partial_\mu} - V(r) - m_q \right] \psi_q(x).
\end{equation}

The normalized quark wave functions $\psi(\vec{r})$ obtained from Eqn (\ref{eq:b}) satisfies the Dirac equation given by
\begin{equation}\label{eq:c}
[\gamma^0 E_q - \vec{\gamma}. \vec{P} - m_q - V (r)]\psi_q (\vec{r}) = 0.
\end{equation}

\begin{table}[b]
\begin{center}
\tabcolsep 0.3pt
    \small
\caption{S-wave $D_s$ ($c\bar{s}$) spectrum (in MeV).} \label{tab2}
\begin{tabular}{|c|c|c|c|c|c|c|}
\hline
  &      &           & \multicolumn{1}{c} {Experiment} && &\\
\cline{4-5}
nL & State &  Present  &  Meson  & Mass \cite{PDG2012} & \cite{Badalian2011} & \cite{Ebert2010}\\
\hline
1S   &$1{^3S_1}$&  2112.3 & $D_s^*$   &  2112.3$\pm$0.5   & - & 2111 \\
     &$1{^1S_0}$&  1970.6 & $D_s$     &  1968.49$\pm$0.32 & - & 1969 \\

2S &$2{^3S_1}$& 2684.4 &  $D_{s1}$(2710) &  $2710_{-7}^{+12}$ & 2728 & 2731 \\
   &$2{^1S_0}$& 2603.9 &        &  2638 \cite{Evdokimov}  & 2656 & 2688 \\

3S  &$3{^3S_1}$& 3195.1 &        &  - & 3200 & 3242 \\
    &$3{^1S_0}$& 3136.9 &        &  - & 3140 & 3219 \\

4S  &$4{^3S_1}$& 3676.0 &        &  - & - & 3669 \\
    &$4{^1S_0}$& 3629.3 &        &  - & - & 3652 \\
\hline
1P   &$1{^3P_2}$&  2572.3 & $D_{s2}$(2573)   &  2571.9$\pm$0.8   & - & 2571 \\
     &$1{^3P_1}$&  2433.7 & $D_{s1}$(2460)   &  2459.6$\pm$0.6   & - & 2574 \\
     &$1{^3P_0}$&  2341.3 & $D_{s0}^*$(2317) &  2317.8$\pm$0.6   & - & 2509 \\
     &$1{^1P_1}$&  2420.4 & $D_{s1}$(2536)   &  2535.12$\pm$0.13 & - & 2536 \\

2P   &$2{^3P_2}$&  3023.2 &                  &   -      & 3045 & 3142 \\
     &$2{^3P_1}$&  2927.7 &                  &    -     & 3020 & 3154 \\
     &$2{^3P_0}$&  2864.1 &                  &     -    & 2970 & 3054 \\
     &$2{^1P_1}$&  2918.4 &  $D_{sJ}$(3040)  &  $3044_{-9}^{+30}$ & 3040 & 3067 \\
   \hline
\end{tabular}
\end{center}
\end{table}
The two component solution of Dirac equation can be written as
\begin{equation}\label{eq:d}
\psi_{nlj}(r) = \left(
    \begin{array}{c}
      \psi_A \\
      \psi_B
    \end{array}
  \right)
\end{equation}
where the positive and negative energy solutions are written as
\begin{equation}\label{eq:e}
\psi_A^{(+)}(\vec{r})=N_{nlj}\left(
    \begin{array}{c}
      \frac{i g(r)}{r} \\
       \frac{(\sigma.\hat{r})f(r)}{r}
    \end{array}
  \right){\cal{Y}}_{ljm}(\hat{r})
\end{equation}
\begin{equation}\label{eq:f}
\psi_B^{(-)}(\vec{r})=N_{nlj} \left(
    \begin{array}{c}
      \frac{i (\sigma.\hat{r})f(r)}{r}\\
      \frac{g(r)}{r}
    \end{array}
  \right)(-1)^{j+m_j-l}{\cal{Y}}_{ljm}(\hat{r})
\end{equation}
and $N_{nlj}$ is the overall normalization constant.
The radial solutions f(r) and g(r) is obtained numerically to yield the energy eigen values. The parameters are fixed to get the ground state masses of $D$ and $D_s$ mesons. The meson radial wave function for $q\bar{q}$ combination is constructed with the respective quark and anti-quark wave functions given by Eqn. (\ref{eq:e}) and \ref{eq:f}. The quark mass parameters  $m_c$, $m_{u,d}$ and $m_s$ are taken as 1.27 GeV, 0.37 GeV and 0.4 GeV respectively.

\section*{Results and Discussion}
The predicted S-wave masses of $D$ and $D_s$ mesons are in very good agreement with experimental \cite{PDG2012} results as given in Table \ref{tab1} and \ref{tab2} respectively. The predicted results of P-wave $D$ meson states, $1^3P_2$ (2455.1 MeV) and $1^3P_0$ (2276.6 MeV) are in good agreement with experimental results of 2462.6 $\pm$ 0.7 MeV and 2318 $\pm$ 29 MeV \cite{PDG2012} respectively. The predicted results of P-wave $D_s$ meson states $1^3P_2$ (2572.3 MeV), $1^3P_1$ (2433.7 MeV) and $1^3P_0$ (2341.3 MeV) are also good found in good agreement with experimental  results $2571.9 \pm 0.8 $ MeV, $2459.6 \pm 0.6 $ MeV and $2317.8 \pm 0.6 $ MeV \cite{PDG2012} respectively.

\section*{Acknowledgments}
  The work is part of Major research project NO. F. 40-457/2011(SR) funded by UGC. One of the author (BP) acknowledges the support through Fast Track project funded by DST (SR/FTP/PS-52/2011).


\end{document}